\documentclass[bibyear]{aa} 
\usepackage{natbib}
\usepackage{multicol}
\usepackage{geometry}
\usepackage{longtable}
\usepackage{caption}
\usepackage{graphicx}
\usepackage{subcaption}
\renewcommand\thesubfigure{\arabic{subfigure}}
\usepackage{txfonts}
\usepackage{hyperref}
\begin{document}

 \title{Unresolved triple systems in the aged open cluster Trumpler~19: Discovery, confirmation, and implications}

 \subtitle{}

 \author{D.~Minniti\inst{1,2}
 \thanks{Based on data from the ESO Public Survey program IDs
 179.B-2002 and 198.B-2004 taken with the VISTA telescope.}
 \and
 R.~K.~Saito\inst{3}, 
 V.~D.~Ivanov\inst{4}, 
 C.~O.~Obasi\inst{1}, 
 M.~G\'omez{\inst1}, 
 P.~W.~Lucas\inst{10}, 
 J.~G. Fern\'andez-Trincado\inst{5}, 
 J.~Alonso-Garc\'ia\inst{6}, 
 E.~R.~Garro\inst{7}, 
 J.~Corral-Santana\inst{7}, 
 A.~Luna\inst{7,16}, 
 Z.~Guo\inst{8, 9}, 
 R.~Kurtev\inst{8}, 
 J.~Borissova\inst{8}, 
 V.~Fermiano\inst{8}, 
 J.~Osses\inst{8}, 
 C.~Morris\inst{8}, 
 B.~Dias\inst{1}, 
 C.~C\'aceres\inst{1}, 
 S.~Saroon\inst{1}, 
 J.~B.~Pullen\inst{1}, 
 P.~Rodriguez\inst{1}, 
 M.~Lad\inst{1}, 
 S.~Federle\inst{1}, 
 I.~Petralia\inst{1}, 
~D.~Bhadrakumar\inst{1}, 
 P.~Esteves\inst{3}, 
 M.~G.~Navarro\inst{11}, 
 T.~Palma\inst{12}, 
 L.~Baravalle\inst{12,17}, 
 D.~Galdeano\inst{13}, 
 M.~V.~Alonso\inst{12,17}, 
 J.~L.~Nilo Castellon\inst{14}
  \and 
  A.~N.~Chen\'e\inst{15}
 }

 \institute{Instituto de Astrof\'isica, Depto. de. F\'isica y Astronom\'ia, Facultad de Ciencias Exactas, Universidad Andres Bello, Av. Fern\'andez Concha 700, Las Condes, Santiago, Chile 
\and Vatican Observatory, V00120 Vatican City State, Italy 
\and Departamento de Física, Universidade Federal de Santa Catarina, Trindade 88040-900, Florianópolis, Brazil
\and European Southern Observatory, Karl Schwarzschildstr 2, 85748 Garching bei München, Germany
\and Centro de investigación en Astronomía, Facultad de Ingeniería, Ciencia y Tecnología, Universidad Bernardo O’Higgins, Av. Viel 1497, Santiago, 8370993, Chile
\and Centro de Astronomía (CITEVA), Universidad de Antofagasta, Av. Angamos 601, Antofagasta, Chile
\and European Southern Observatory, Casilla 19001, Vitacura, Santiago, Chile
\and Instituto de F\'isica y Astronom\'ia, Universidad de Valpara\'iso, Av. Gran Breta\~na 1111, Playa Ancha, Casilla 5030, Chile
\and Chinese Academy of Sciences South America Center for Astronomy (CASSACA), National Astronomical Observatories, CAS, Beijing 100101, China
\and Centre for Astrophysics Research, University of Hertfordshire, College Lane, Hatfield AL1 09A, UK
\and Istituto Nazionale di Astrofisica (INAF), Rome Astronomical Observatory, Via Frascati, 33, 00078 Monte Porzio Catone, RM, Italy
\and Observatorio Astronómico de Córdoba, Universidad Nacional de Córdoba, Laprida 854, X5000BGR, Córdoba, Argentina
\and Departamento de Geofísica y Astronomía, CONICET, Facultad de Ciencias Exactas, Físicas y Naturales, Universidad Nacional de San Juan, Av. Ignacio de la Roza 590 (O), J5402DCS, Rivadavia, San Juan, Argentina
\and Departamento de Astronomía, Universidad de La Serena, La Serena, Chile
\and NSF NOIRLab, 670 N. A'ohoku Pl. Hilo 96720, Hawaii, USA 
\and Istituto Nazionale di Astrofisica (INAF), Osservatorio Astronomico di Capodimonte, Salita Moiariello 16, 80131, Naples, Italy
\and Instituto de Astronom\'{\i}a Te\'orica y Experimental, (IATE-CONICET), Laprida 854, X5000BGR, C\'ordoba, Argentina
}
\offprints{vvvdante@gmail.com}

\date{Received June xx, 2026; accepted XXXX xx, 2026}

 
 \abstract
 { It is difficult to  identify unresolved triple and
   higher-order systems  beyond  the solar  vicinity.  Star  clusters
   yield  controlled  environments  with  known  physical  parameters,
   providing a way to explore homogeneous stellar samples.  }
 { Combining modern  optical and near-IR surveys, our main  goal is to
   explore the  region of  the color-magnitude diagrams  located above
   the  sequence  of  equal-mass   binaries,  which  is  inhabited  by
   unresolved triple and quadruple systems.  Specifically, here we aim
   to discover  and characterize the  presence of multiple  systems in
   the aged open star cluster Trumpler~19.  }
 {We  selected  cluster members  using  Gaia  astrometry  and
   photometry from  the VVVX and DECAPS2 surveys. We made optical  and  near-IR color-magnitude  diagrams to  select
   stars  located   $>0.75$~mag  above  the  cluster   main sequence,
   including suitable unresolved triple and quadruple systems.  }
 { We  confirm previous results  that there  is a sizable  sequence of
   unresolved binaries in  Trumpler~19.  We also  report the discovery
   of a  sequence of unresolved  multiple systems  in this
   cluster, and confirm some of these multiple systems via radial velocities measured  by the  ESO KMOS  VVVX-GalCen
   spectroscopic  survey.   The fraction  of  multiple  stars in  this
   cluster is comparable  to that of the solar  neighborhood, with the
   caveat  that  the   local  sample  probes  deeper   down  the  main sequence. 
  The fraction of unresolved triples and quadruples is also
   three times higher than  that of the open cluster M~67, which is of a similar age and
   chemical composition.   
    }
 { The discovery of unresolved  triple systems in Trumpler~19 leads to
   our main conclusion that these  systems were able to survive within
   the cluster for  nearly 4~Gyr.  In addition,  this cluster contains
   more  multiple systems  than M~67,  which has  a similar  total
   present mass.  Further investigations are  needed to unveil if this
   is a primordial feature or the result of dynamical evolution.  }


 \maketitle
%

\section{Introduction}\label{Introduction}

Multiple stars  are common  in the  solar neighborhood,  with binaries
being  the majority,  although triple  and higher-order  multiples are
also present \citep{Duquennoy1991,Henry2018,Reyle2021}.   The study of
multiple  systems is  very  demanding  in telescope  time,  but it  is
important for understanding  a variety of astrophysical  problems, such as
star   formation,  census,   mass  budget,   hierarchy,  architecture,
composition,  and dynamical  evolution \citep[e.g.,][ and references
  therein]{Tokovinin2023,Rappaport2024,Vrijmoet2026}.   It  has   been
difficult to  identify and characterize these  triple and higher-order
systems beyond the solar vicinity because they become unresolved.

Star cluster  studies rely on  the fitting  of isochrones to  the main sequence and giant branch, using mean ridge lines. The binary sequence
on top of the main sequence (up  to $0.75$~mag brighter for equal-mass binaries)
is       generally      taken       into      account       \citep[see
][]{Jadhav2021,Donada2023,Jiang2024}, but stars  that are located well
above these  sequences are  often discarded  as field  stars. However,
unresolved triples and  quadruples should be located  above the binary
sequence and  cannot be neglected  a  priori. In this  work, we
report the first  identification of a sequence  of unresolved multiple
stellar  systems,  from triples  to  higher-order  systems, above  the
binary sequence of an aged open cluster.

This study is a first  step toward investigating triple and quadruple
stellar systems  within an open  cluster in detail.  The  advantage of
such a  controlled environment  is that we  know the  ages, distances,
reddenings, and so on, as  these are  provided by  the measured  mean open
cluster physical parameters. The clusters also  offer a way to study a
homogeneous large  sample of stars.   In turn, the  masses, gravities,
and effective  temperatures of  the multiple  systems can  be derived
with the aid of the cluster isochrones. Also, spectroscopic orbits can
be obtained in some  cases using follow-up observations. Time-resolved
spectroscopy is particularly desirable in  order to obtain the orbital
parameters for some of these complex systems. Of course, this can only
be achieved if we can secure cluster membership, which is now possible
thanks to the data release 3  of the Gaia mission, the VISTA Variables
in the  Vía Láctea  eXtended (VVVX)  ESO public  survey, and  the Dark
Energy Camera Plane Survey 2 (DECaPS2).

The subject  of the present  study is Trumpler~19 (FSR1565 or
Teutsch106), which  is a well-known open  cluster with age $t=3.7\pm0.3$~Gyr
, located  in  the  Galactic plane  at  $l, b=  290.1849,
+02.8847$~deg,  at  a  distance of $D=2.3$~kpc  and  moderately  reddened
\citep[$A_V=0.77$~mag; ][]{Sheikh2025}. This open cluster is contained
in the  tile e0808 of  the VVVX  survey, which provides  multicolor
near-IR imaging ($JHK_{\rm s}$), and multi-epoch photometry from years
2017 to  2023 in the $K_{\rm  s}$ band \citep{Saito2024}.  Trumpler~19
represents  a controlled  volume-limited sample,  with known  physical
parameters.  Its galactocentric distance is $R_G=7.8$~kpc,  and it is
located     at     $z=146$~pc     above     the     Galactic     plane
\citep{CantatGaudin2020,Dias2021}, giving a similar environment as our
Sun.  Its velocity  and PMs define a nearly circular  orbit within the
plane,   although  eccentric   enough  to   cover  between   peri  and
apogalacticon      \citep[$R_{peri}=6.6$~kpc,      $R_{apo}=11.0$~kpc,
][]{Sheikh2025}.

Unresolved  binaries should  be  brighter than  single  stars and  are
therefore located above the main sequence. Their brightness depends on
the  mass  ratio, $q$,  which reaches a maximum of $1.0$,  corresponding  to
equal-mass binaries. Therefore, this sequence of unresolved equal-mass
binaries  should  be $0.75$~mag  brighter  than  the MS.   Conversely,
unresolved equal-mass triplets should  be $1.19$~mag brighter than the
MS and  $0.44$~mag brighter  than the binaries;  unresolved equal-mass
quadruples should  be $1.51$~mag brighter  than the MS  and $0.32$~mag
brighter than the triples; and unresolved equal-mass quintuples should
be $1.75$~mag  brighter than the  MS and $0.24$~mag brighter  than the
quadruples, and  so on  for higher-order  multiple systems.  Note that
these observables should be independent of distance and reddening.
Considering these limits, we selected 390 candidate unresolved binaries, 
and 25 candidate  triple and quadruple systems.

However, we must consider other alternative explanations, as there are
other ways  to place  stars above  the cluster  MS: through
photometric   scatter,  through   variability,  through   blending  or
mismatches, through  a shorter distance,  through reddening,  or through
the presence of  circumstellar disks.  We analyzed  these hypotheses in
turn,  in  order  to  show   that  they  can  be  readily  eliminated.
Photometric errors  were discarded because these  magnitude differences
between the  sequences are much  larger than the  measured photometric
errors ($\sigma < 0.02$~mag in all optical and near-IR passbands), and
also because  we used  a variety  of optical and  near-IR CMDs  and the
multiple systems are brighter in all passbands.  Variable stars can be
readily  identified thanks  to the  multi-epoch observations  of Gaia,
VVVX,  and  WISE, and  we  find  that, even  though  there  are a  few
interesting  variables, they  cannot  account for  all the  unresolved
multiple  system candidates.   Blending and  mismatches with  a bright
field  source (either  foreground or  background) were  discarded using
Gaia PMs  and parallaxes.   The distance vector  moves objects  up and
down vertically in  the CMDs, but in this case the shorter distances
were  discarded  using Gaia  parallaxes. The  reddening vector  makes
objects  fainter  and  redder,   but  differential  reddening  can  be
discarded using  the relatively  homogeneous spatial  distribution and
the tightness  of the  cluster sequences  in the optical and near-IR color-color diagrams.
Finally, the
contribution  from circumstellar  disks is  important in  the case  of
young objects, as they contribute to  IR excess in the spectral energy
distribution, but these disks should not be present in stars belonging
to clusters as old as Trumpler~19.  In spite of the thorough selection
discussed  above,   we  considered  it  necessary   to  obtain  radial
velocities in order to have an independent membership confirmation
 for the sample of unresolved multiple systems detected here.

\section{Optical and near-IR photometry}\label{Selection}

\begin{figure}[!]
\centering
\includegraphics[width=80mm]{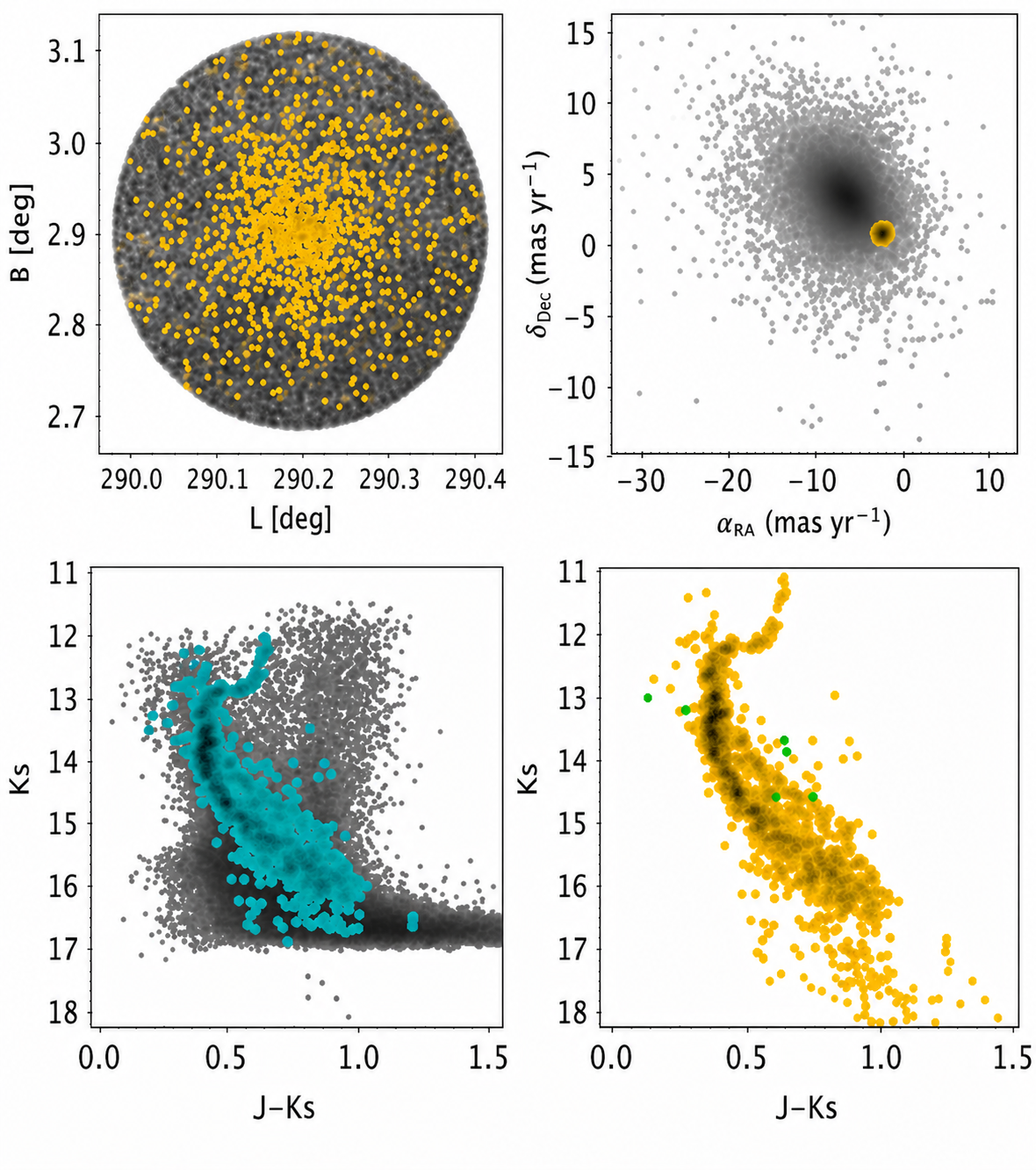}
\caption{ Spatial, proper-motion, and colour--magnitude distributions of Trumpler 19. The  top   panels  show  the  spatial   and  proper  motion
  distributions  for Trumpler~19  cluster members  (yellow) and  field
  stars (gray).  The bottom left panel  shows the VVVX near-IR CMD for
  field  and   cluster  stars   (in  gray  and   cyan,  respectively),
  illustrating  the large  field contamination.   Stars brighter  than
  $K_{\rm s}=12$~mag are saturated in the VVVX photometry.  The bottom
  right panel highlights the Gaia DR3 variable stars in green, located
  in the blue straggler star region and above the MS.  }
\label{fig:figura1}
\end{figure}

\begin{figure*}[!]
\centering
\includegraphics[width=172mm]{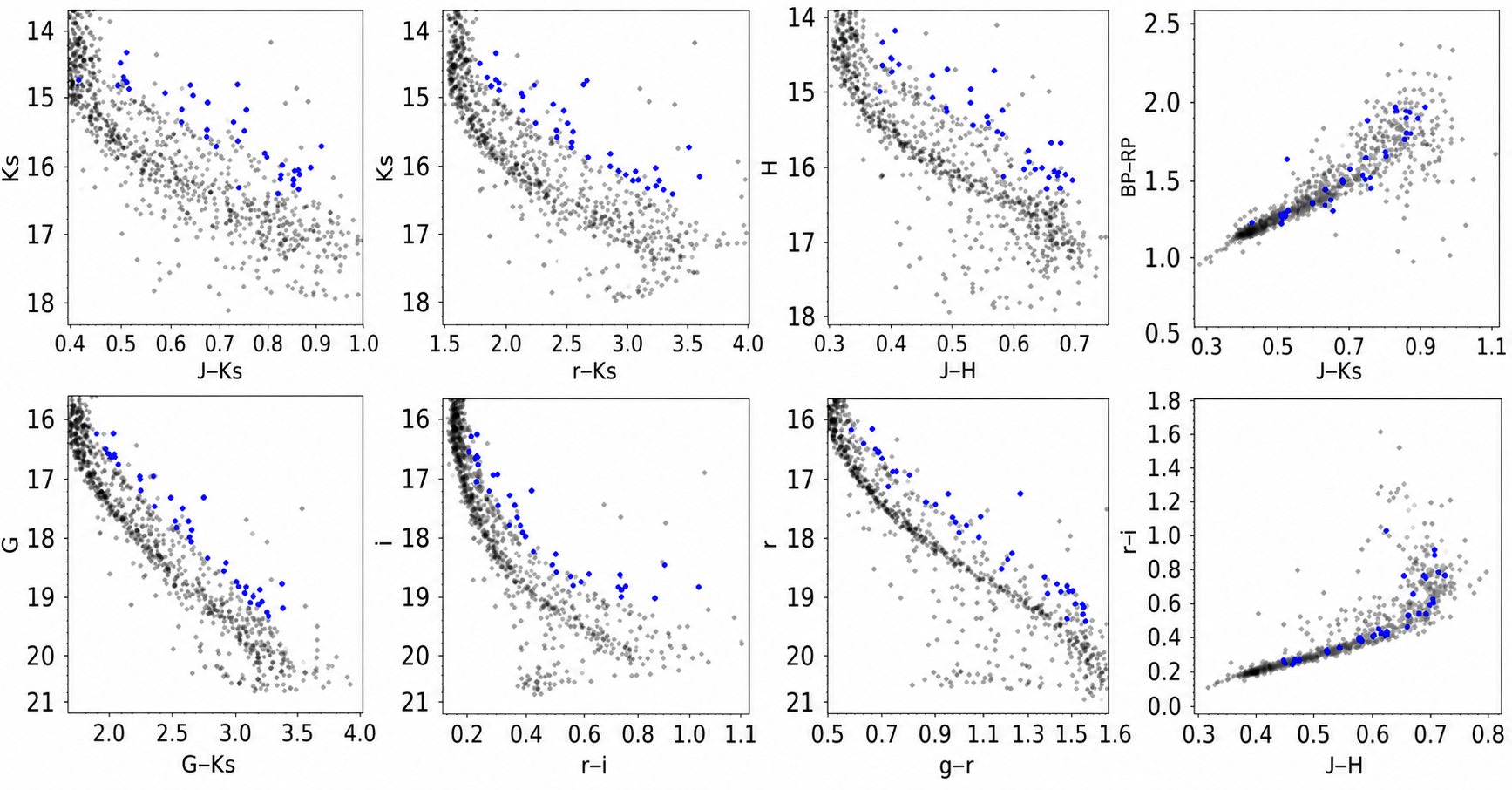}
\caption{Optical  and near-IR  CMDs  and color-color  diagrams for  PM-selected  members of  Trumpler~19 (gray  circles), highlighting  the
  selected unresolved  triple and quadruple candidates  (blue points).
  These  are located  above the  sequence  of equal-mass binaries  in
  general.  }
\label{fig:figura2}
\end{figure*}

\begin{table*}
 \caption{Trumpler~19 candidate unresolved systems.
}
\label{tab:data}
\centering
\scriptsize
\begin{tabular}{l c c c c c c c c c c c c}
\hline
\hline
\textbf{Gaia DR3 ID} & 
\textbf{$G$} & 
\textbf{$BP-RP$}  & 
\textbf{$J$} & 
\textbf{$H$} &
\textbf{$K_{\rm s}$}&
\textbf{$g$} &
\textbf{$r$} &
\textbf{$i$} &
\textbf{$z$} &
\textbf{$Y$} &
\textbf{Prob} &
\textbf{$R_V$}\\
&
 \textbf{(mag)} &
 \textbf{(mag)} &
 \textbf{(mag)} &
 \textbf{(mag)} &
 \textbf{(mag)} &
 \textbf{(mag)} &
 \textbf{(mag)} &
 \textbf{(mag)} &
 \textbf{(mag)} &
 \textbf{(mag)} &
 \textbf{} &
 \textbf{(km s$^{-1}$)}\\
\hline
  5339935386069088384 & 16.75 & 1.21 & 15.192 & 14.769 & 14.621 & 17.37 & 16.61 & 16.36 & 16.23 & 16.17 & -- & $-24.68 \pm 0.41$\\
  5339935351716750208 & 17.17 & 1.16 & 15.608 & 15.194 & 15.041 & 17.82 & 17.05 & 16.80 & 16.66 & 16.59 & 0.999 & $-33.04 \pm 0.48$\\
  5339935416105335424 & 18.77 & 1.73 & 16.671 & 16.068 & 15.887 & 19.92 & 18.67 & 18.17 & 17.92 & 17.80 & 0.293 & $-20.82 \pm 12.03$\\
  5339935145550936576 & 17.08 & 1.17 & 15.549 & 15.133 & 14.989 & 17.71 & 16.96 & 16.72 & 16.59 & 16.53 & 0.994 & $-31.95 \pm 1.38$\\
  5339935420428856576 & 17.97 & 1.23 & 16.263 & 15.835 & 15.647 & 18.61 & 17.80 & 17.50 & 17.34 & 17.27 & 0.974 & $-17.18 \pm 2.44$\\
  5339936176343091840 & 16.91 & 1.09 & 15.432 & 15.034 & 14.885 & 17.53 & 16.80 & 16.57 & 17.86 & 16.83 & 0.998 & $-34.34 \pm 7.42$\\
  5339936038904090240 & 16.51 & 1.06 & 15.123 & 14.760 & 14.618 & 17.05 & 16.40 & 16.20 & 16.10 & 16.05 & 0.999 & $-21.55 \pm 0.58$\\
  5339940986706705152 & 17.25 & 1.23 & 15.631 & 15.167 & 15.002 & 17.90 & 17.15 & 16.87 & 16.71 & 16.64 & 0.001 & $-21.65 \pm 0.51$\\
  5339934801953460736 & 17.42 & 1.26 & 15.761 & 15.273 & 15.141 & 18.13 & 17.32 & 17.01 & 16.86 & 16.77 & 0.993 & $-28.41 \pm 1.57$\\
  5339940952346951168 & 16.95 & 1.16 & 15.474 & 15.038 & 14.924 & 17.54 & 16.84 & 16.61 & 16.49 & 16.42 & 0.999 & $-11.61 \pm 1.23$\\
  5339940982396345088 & 18.50 & 1.69 & 16.491 & 15.858 & 15.685 & 19.61 & 18.37 & 17.94 & 17.70 & 17.60 & 0.991 & $-13.82 \pm 0.60$\\
  5339941360341104768 & 17.80 & 1.38 & 16.040 & 15.524 & 15.386 & 18.62 & 17.66 & 17.34 & 17.16 & 17.08 & 0.873 & $-23.09 \pm 2.34$\\
  5339935626587165312 & 18.47 & 1.57 & 16.497 & 15.902 & 15.717 & 19.48 & 18.36 & 17.92 & 17.68 & 17.59 & 0.952 & $-36.12 \pm 2.53$\\
{\bf 5339942081895605888}  & 17.40 & 1.30 & --         & 15.215 & 15.011 & 18.22 & 17.36 & 17.11 & 16.93 & 16.83 & 0.997 & $-28.52 \pm 0.57$\\
  5339940982376753408 & 17.66 & 1.33 & 15.969 & 15.446 & 15.356 & 18.40 & 17.50 & 17.19 & 17.02 & 16.95 & 0.917 & $-19.28 \pm 2.32$\\
  5339941433383341440 & 16.62 & 1.12 & 15.172 & 14.780 & 14.662 & 17.20 & 16.51 & 16.29 & 16.18 & 16.12 & 0.999 & $-19.68 \pm 1.84$\\
  5339935008119347072 & 17.58 & 1.33 & 15.913 & 15.414 & 15.272 & 18.35 & 17.45 & 17.15 & 16.99 & 16.91 & 0.997 & $-32.73 \pm 4.77$\\
  5339894051274700288 & 18.59 & 1.72 & 16.428 & 15.755 & 15.627 & 19.71 & 18.49 & 17.98 & 17.68 & 17.57 & 0.995 & $-56.84 \pm 0.56$\\
  5339894188749205376 & 17.04 & 1.12 & 15.568 & 15.126 & 15.026 & 17.63 & 16.93 & 16.70 & 16.58 & 16.52 & 0.005 & $-31.60 \pm 0.32$\\
{\bf 5339941502102854272}  & 16.97 & 1.32 & --          & 14.784 & 14.601 & 17.72 & 16.86 & 16.57 & 16.38 & 16.30 & 0.990 & $-20.53 \pm 1.13$\\
  5339894085669964288 & 16.55 & 1.05 & 14.997 & 14.888 & 14.584 & 16.80 & 17.25 & 17.23 & 16.84 & 16.58 & -- & $-42.42 \pm 8.01$\\
{\bf   5339934973752164096}  & 17.04 & 1.18 & 15.452 & 15.081 & 14.80  & 17.78 & 16.95 & 16.71 & 16.56 & 16.49 & 0.999 & $-105.58:$\\
\hline
\end{tabular}
\tablebib{The targets highlighted in bold are classified as variable stars in the Gaia DR3 database.}
\end{table*}

\begin{figure}[!]
\centering
\includegraphics[width=80mm]{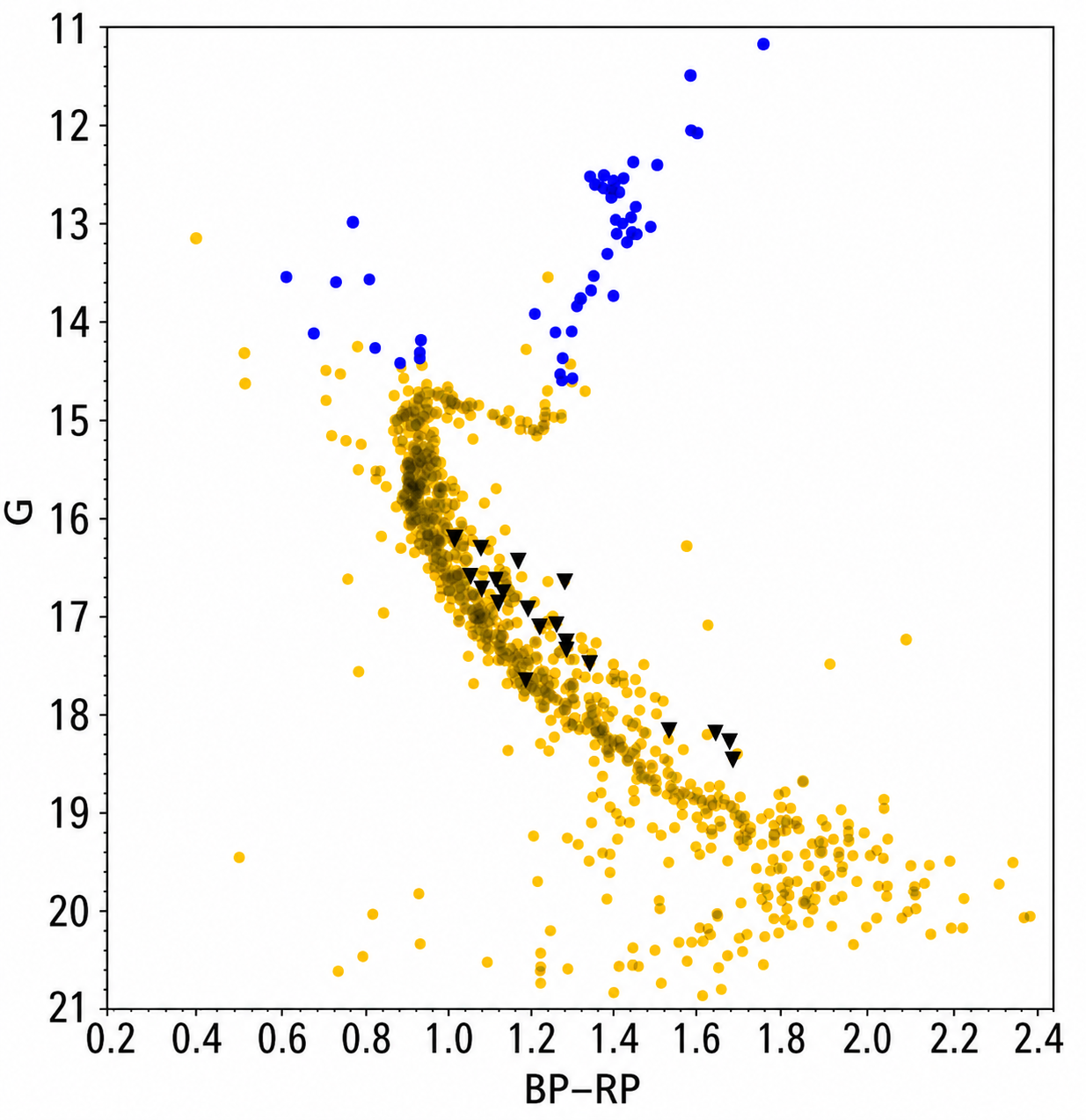}
\caption{  Gaia DR3  CMD  for Trumpler~19  (yellow), highlighting  the
  stars with measured  Gaia radial velocities (blue  circles), and the
  candidate   unresolved  triples   with   KMOS  spectroscopy   (black
  triangles).  It  is clear  that KMOS targets  are much  fainter, and
  located well above the cluster MS.  }
\label{fig:figura3}
\end{figure}

There are significant  impediments to the studies of  open clusters in
the  Galactic plane.  In particular,  the membership  determination is
complicated due to differential reddening and field contamination. The
various datasets  are affected by different  observational biases, and
completeness is always  an issue. In consequence,  most studies prefer
to be conservative,  discarding outliers, such as stars  away from the
known color-magnitude  diagram (CMD) sequences. Triple  systems may be
missed when  using the mean  ridgelines to  fit the CMD  sequences. In
this context, the sequence of  unresolved binaries has been identified
because it  is well populated in  some clusters, but the  more complex
unresolved  systems (triples,  quadruples, and  beyond) have  not been
clearly  seen before.  Here  we  report the  discovery  of the  triple
sequence in the open cluster Trumpler~19.

As a first step, we used the  astrometry from Gaia DR3 to select proper
motion  members of  the  Trumpler~19, (see  Figure~\ref{fig:figura1}),
considering  objects with  $1.0$~mas\,yr$^{-1}$  of  the mean  cluster
motion,  taken  to  be  $PM_{RA}  =  -1.64  \pm  0.13$~mas\,yr$^{-1}$,
$PM_{DEC}  =  -1.20  \pm 0.17$~mas\,yr$^{-1}$  \citep{Sheikh2025}.   A
selection refinement is given by the parallaxes, for which we adopted an
interval   in   parallax   of    $0.365   \lesssim   \varpi   \lesssim
0.515~\mathrm{mas}$.  We  extracted the near-IR point spread function (PSF)  catalogs from the
VVVX survey  in the $JHK_{\rm s}$ passbands  from \citet{AlonsoGarcia2026}, and
the   $grizY$   PSF   photometry   from   the   DECAPS   database   from
\citet{Saydjari2023}. The matches  between Gaia,  VVVX, and  DECaPS2
required a positional difference of $<0.6^{\prime\prime}$, but it should be noted that this
upper limit  does  not  matter  much  because  the coordinates  match  within
$0.1^{\prime\prime}$.   In  summary,  we  find  $N=1258$~members  with
Gaia+VVVX+DECaPS2 photometry, including ten Gaia variable stars. These
are out  of $N=55826$~stars within  15~arcmin in the Gaia  DR3 catalog
and  out of  $N=104064$~stars  within 15~arcmin  in  the VVVX  catalog
(i.e.,  only   about  1\%  of   the  available  near-IR   sources  were
preselected as cluster members for this study). Therefore, because of
the availability  of quality photometry and  astrometry from different
surveys, we were able to select  a well-distilled sample of Trumpler~19
members.  
The membership probability values listed in Table 1 were computed with a Gaussian mixture model (GMM) fit to positions, proper motions, and parallaxes following previous works (Dias et al. 2021, Hunt \& Reffert 2023, Sheikh et al. 2025). The PM spread of members is quite small ($\sigma \approx 0.1$ mas/yr in PMRA and PMDEC) after removing sources with PM uncertainties of $>1$ mas/yr, parallax uncertainties $> 1$ mas, or a renormalised unit weight error (RUWE) of $>1.4$ prior to the fit. We note that probability values are not given for two of our targets that did not comply with these stringent criteria because they have higher RUWEs (Gaia DR3 5339935386069088384 with $RUWE=1.495$, and Gaia DR3 5339894085669964288 with $RUWE=1.816$). Also note that a couple of the candidates have matching velocities but low GMM membership probabilities.
The cluster is very near to the Galactic plane, and after using proper motion and parallax selections some field stars may still remain. The contamination was evaluated using two comparison fields of similar areas located at 1 degree away from the cluster, at the same Galactic latitude to the east and to the west, sufficiently far away not to include cluster members but close enough to sample the same Galactic disk populations. The average contamination limit is estimated to be 6 to 10\% for the stars selected using identical cuts to the cluster stars. This estimate agrees with the proportion of RV nonmembers measured by KMOS, and therefore field contamination can be readily eliminated as a major concern.

In  addition, we match the  sources detected by the Wide Field Infrared Survey Explorer (WISE) in the
cluster  field,  finding  $N=175$~members  matched  with  mid-IR  WISE
photometry in the W1, W2, W3,  and W4 bands.  These include three WISE
variable stars, two of which are  located in the blue straggler region
of the  CMDs and one of which is located  well above the MS.   The WISE photometry
covers   the  red   giant  branch   and  the   turnoff  region,   but
unfortunately it does not reach deep down the MS.

We then made a variety of optical and near-IR CMDs and color-color diagrams, where the
main sequence    of    Trumpler~19    is    very    well-defined
(Figure~\ref{fig:figura2}).   In  these  CMDs,  we  identified  multiple
sources     located     above     the    cluster     main sequence,
the majority of which should be unresolved
binaries, in agreement with previous works \citep[e.g., ][ determine a
  binary fraction of $0.66$ for this cluster]{Sheikh2025}. However, we
also detect a  number of objects that are  significantly brighter than
the binary sequence, which we propose to be unresolved multiple systems
(mostly triples and quadruples).

Multiple  members located  near  the turnoff  region  are in  general
difficult to interpret.   We selected a region of the  MS well below the
turnoff  where  the  selection  of  unresolved  multiples  should  be
cleaner, because they  would be better separated,  forming a different
sequence on  their own. The  Trumpler~19 sample studied  here consists
mostly of late G and K-type  dwarfs, which are fainter than the cluster
turnoff.   The   color  cuts  applied   to  select  the   sample  were
$1.0<BP-RP<1.8$~mag,   $1.75<G-K_{\rm    s}<3.5$~mag, and $0.45<J-K_{\rm
  s}<0.85$~mag (the projection of these  cuts onto other colors can be
seen in Figure~\ref{fig:figura2}).  Taking  into account the reddening
values   for  Trumpler~19   corresponding   to  $A_V=0.77$~mag   (i.e.,
$E_{BP-RP}=0.35$ mag, $E_{J-Ks}=0.12$ mag, and $E_{G-K{\rm s}}=0.56$~mag),
this color  selection largely avoids  the early G-type stars  and also
the sequence of M-type stars.

\begin{figure*}[!]
\centering
\includegraphics[width=142mm]{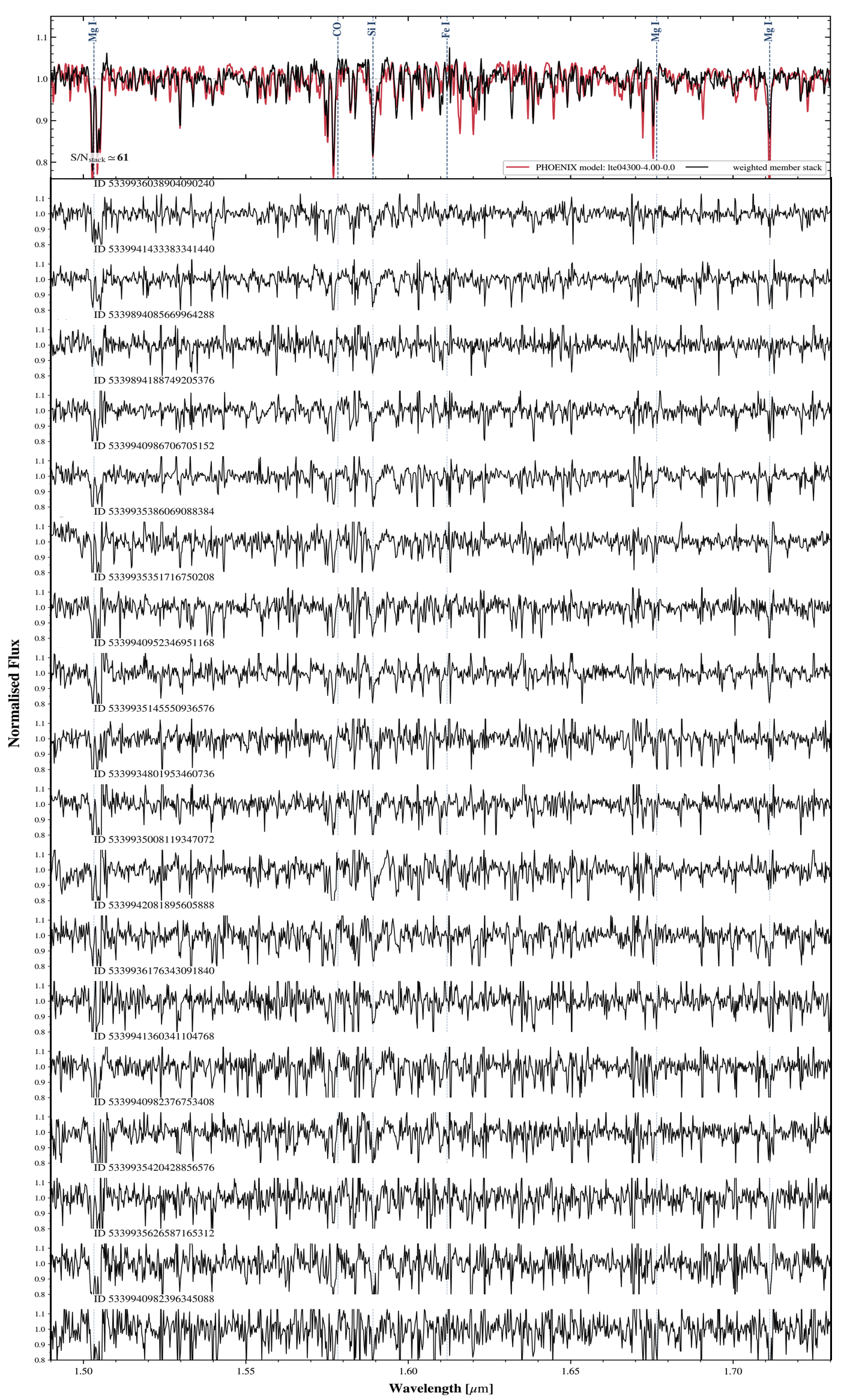}
\caption{  
 H-band KMOS spectra ($R=4000$) for the candidate unresolved multiple systems in Trumpler~19 (see Table 1).
These individual spectra have been normalized and displaced vertically, and some strong spectral features are labeled. 
Increased scatter is evident in regions where residuals from the telluric line corrections are present.
The top panel highlights the stacked  weighted spectrum (total $S/N=61$) compared with a PHOENIX LTE model from Husser et al. (2013) for $T_{eff}=4300$ K, $log~g = 4.0$ and solar metallicity (red line).
}
\label{fig:figura4}
\end{figure*}

\begin{figure}[!]
\centering
\includegraphics[width=84mm]{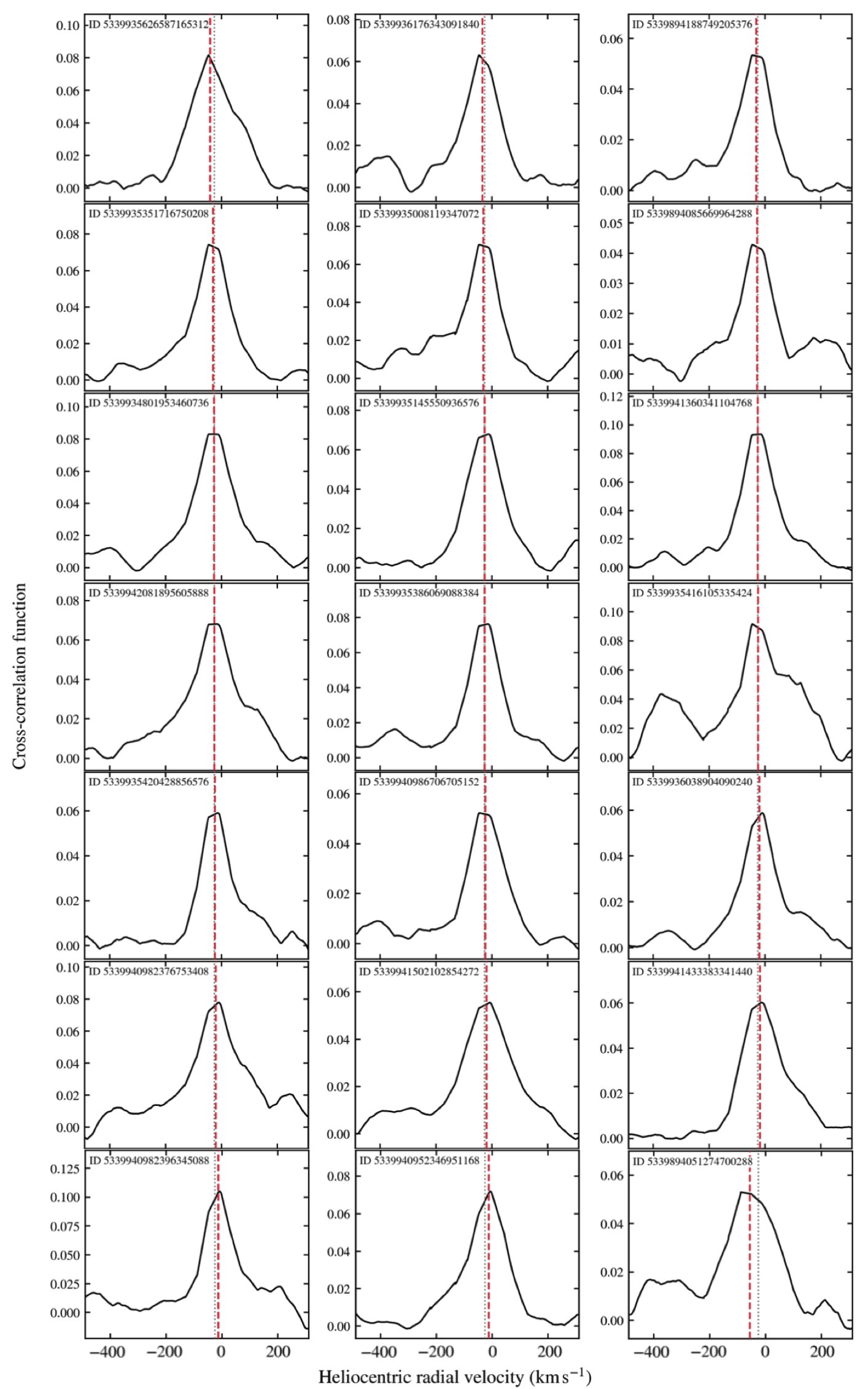}
\caption{  
Peaks of the CCFs measured from the  KMOS spectra of Trumpler~19 members listed in Table 1.
The vertical lines indicate the measured radial velocity (dashed red line) compared with the mean cluster velocity (dotted black line).
}
\label{fig:figura5}
\end{figure}

\section{The KMOS spectroscopy }

The  near-IR  spectroscopy  was performed  with  the  KMOS
spectrograph  mounted   at  the  VLT   UT1  located  at   ESO  Paranal
Observatory,  as part  of the  ESO Large  Spectroscopic Public  Survey
VVVX-GalCen (G\'omez et  al.  2026,   in
  prep.). We selected  18 triple candidates plus 4  binaries 
(Figure~\ref{fig:figura3} shows  the selected  stars in  context). The
selection among  the candidate targets  was made in order  to optimize
the  KMOS  arm  allocation,  and  the  targets  with  measured  radial
velocities are  located well within the  core of the cluster  owing to
the  KMOS  $7.5^{\prime}$ field  of  view.   We acquired  the  near-IR
spectra with  a single  KMOS pointing  in the  $H$-band configuration,
using an exposure time of 60~min  starting at UT~00:47:37 on the night
of 10  May 2026, under $0.6^{\prime\prime}$~seeing.   The spectra were
reduced and calibrated  using the  ESO Data Processing System  (EDPS; Freudling et al. 2004) with the dedicated KMOS pipeline, resulting in
22  good spectra with  $10 < S/N < 22$ (2 arms  were broken).
Figure 4 shows the normalized individual KMOS spectra, and Figure~\ref{fig:figura5} shows the respective cross correlation functions. These are 22 good spectra arbitrarily displaced vertically to better show the spectral lines and where the noisier portions reveal the regions telluric subtraction remains imperfect. No obvious multiple-line systems or wide-line systems are firmly detected at the KMOS resolution. We estimate that wide systems with velocity amplitudes smaller than $\sim 100$ km/s would remain undetected for our single epoch spectra. Multiple epochs would be needed to confirm the spectroscopic binary or triple nature of these systems.

\begin{figure}[!]
\centering
\includegraphics[width=80mm]{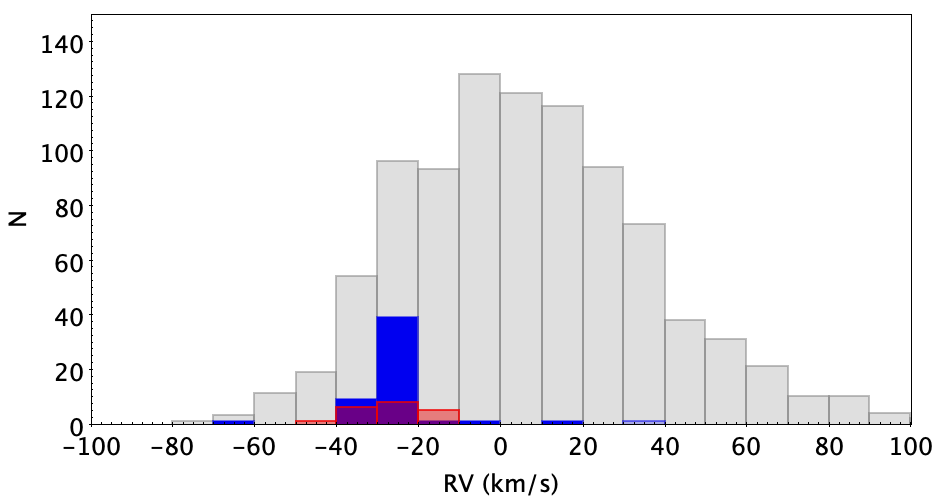}
\caption{ $R_V$ distribution  for Trumpler~19 from Gaia  (blue) and the
  KMOS  targets  (red) compared  with  the  field stars  (gray).   The
  tighter radial velocity concentration of the cluster stars with respect to the field stars is evident.  }
\label{fig:figura6}
\end{figure}

The radial  velocities were  measured using
cross correlation,  and calibrated  applying a zero  point correction
based on  the standard globular cluster  M22. We note that  due to the
moderate instrumental resolution  ($R = 4000$), we  would be unable
to discriminate  most spectroscopic  binaries or to  measure rotations
from  the line  profiles, but  it  is possible  to measure  velocities 
accurate  to  a few kilometres per second in  order to determine cluster
membership. Figure~\ref{fig:figura6}   shows  the   measured  radial
velocity distribution, which  exhibits a small dispersion  ($\sigma = 6
\pm 1$~km~s$^{-1}$) that is similar to other clusters observed for our
VVVX-GalCen spectroscopic survey, from which we conclude  that all but two stars are
confirmed members of the cluster. We  measure a mean velocity for this
cluster of $R_V= -25.65  \pm 1.75$~km~s$^{-1}$, in good  agreement with the
Gaia    mean   velocity  of  $R_V   =    -26.3   \pm    1.7$~km~s$^{-1}$
\citep{Drimmel2023}.   Table~1  lists   the  measured  velocities  for
Trumpler~19 members  observed,  including the  two stars  that are
considered nonmembers.   Particularly interesting is the  case of the
$K_{\rm s}=  14.60$~mag star Gaia DR3  5339941502102854272, located at
$RA= 168.676460$~deg,  $Dec= -57.523124$~deg, which is  variable in the
WISE catalog but not  flagged as a variable by Gaia.   This target
complies with  all of the selection  cuts, and it can  be considered a
cluster  member  according   to  the  $R_V=-20.53  \pm1.13$~km~s$^{-1}$
(spectrum $S/N=19$).

\section{Results and discussion}

There  are  triple systems  known  in  open  clusters that  have  been
previously well  characterized. We  can cite  a few  specific examples
such as  M~67 \citep{Sandquist2003},  NGC2516  and  NGC6530
\citep{Gonzalez2014},  NGC6819  \citep{Brewer2016},  and  Ruprecht~147
\citep{Torres2021}. However, a sequence such as the one we observe here has never
been reported.
The  comparison  with the  open  cluster  M~67  in particular  is  very
interesting, because this  cluster has a similar  age ($t=4$~Gyr), and
metallicity  ($Z=0$), as  Trumpler~19,  and it  is  very well  studied
because  it  is  located  at  $d=900$~pc,  significantly  closer  than
Trumpler~19,  with $d=2300$  pc.  The CMDs of M~67 \citep{Leiner2025} show that there are a couple
of stars in the region associated with triple and quadruple systems,
but these do not form a sequence similar to the one we find in
Trumpler~19. This
indicates that  the multiplicity fractions  for these two  clusters of a
similar age are very different.  
We measure the unresolved triple and higher-order fraction as defined by
\citet{Offner2023}, finding that this fraction in Trumpler~19
($\mathrm{THF} = 0.084 \pm 0.010$) is higher than that of the open
cluster M~67 ($\mathrm{THF} = 0.029 \pm 0.010$). Therefore, there are nearly three times more unresolved triples and quadruples in Trumpler19 than in M~67.
This fraction was measured within the color range $2.25<BP-Ks<4.75$ mag for Trumpler 19, and $1.80<BP-Ks<4.30$ mag for M~67, respectively (the color difference is due to the different mean cluster reddenings). This color range corresponds approximately to MS stars of spectral types roughly from G5V to M0V, temperatures from 3800 K to 5300 K, and masses within $0.5 < M/Mo < 0.9$ (even though it is not straightforward to apply these parameters for our targets because they are composite systems).
Somehow Trumpler~19 managed to create
more multiple systems, or M~67  has more efficiently destroyed the multiple
systems.  Note that  M~67 is an  evolved cluster that has a comparable
present mass of $M=2000$ $M_{\odot}$ and that has  lost most of its  mass in
the past \citep{Hurley2005}, similar to Trumpler~19,  which has a present mass
$M=2400$   $M_{\odot}$  and which has  also   lost  most   of  its   mass
\citep{Sheikh2025}.

It  is  well  known that  a  large  fraction  of  stars in  the  solar
neighborhood    appear    in    binary   or    higher-order    systems
\citep{Duquennoy1991}.
For example,  \citep{Tokovinin2014,Tokovinin2023}  found that the  frequency of field
multiple systems follows the distribution $N=1:2:3:4:5$, with relative
numbers of $f=54:33:8:4:1$..This means that, on average, every 54
isolated stars there  are 111 stars in multiple systems. 
This is in good agreement with  the results from \citet{Raghavan2010}  
that reported 56\% singles, 33\% binaries, 8\% triplets, and 3\% higher multiplets.  
Also, more recently, the updated sample from  \cite{Reyle2021} in the 10~pc solar
neighborhood yields $N=246$  single stars, 69 binaries,  19 triples, 3
quadruples, and  2 quintuples  (this means 339  objects in  total that
would be  unresolved if placed  at the distance of  Trumpler~19). That
yields 217 stars belonging to multiplets  (47\% of the total stars, or
64\%  of   unresolved  sources  at  $2.3$~kpc).    The  RECONS  survey
\citep{Henry2018} gives similar numbers, with 262 singles, 66 doubles,
14 triples, 3  quadruples, and 2 quintuples out of  462 stars in total
(347 unresolved objects if placed at  $2.3$~kpc).  We can adopt a mock
stellar  neighborhood 10pc  sample placed  at $2.3$~kpc  that contains
$N=343$   objects,    including   unresolved    $N_{bin}=68$   (20\%),
$N_{tri}=17$  (5\%),  $N_{quad}=3$  (1\%), and $N_{quint}$=1  (0.3\%).  If Trumpler~19  
maintains this  mock  proportion,  it
should  have:  $N_{bin}=249$~binaries,  smaller  than
measured  $N_{bin}=390$,  and  $N_{tri}=62$  triples,
higher than measured $N_{tri}=25$.   Thus, it appears that Trumpler~19
has  proportionally more  binaries  and fewer  triples  than the  solar
neighborhood.   The caveat  is that  we are  comparing different  mass
ranges  because  the nearby  sample  is  dominated by  fainter  stars.
While the comparison with M~67 is straightforward because these clusters have similar ages and metallicities, the comparison with the field stars is more difficult, and therefore uncertain. To summarize, the caveats about the comparison of Trumpler 19 with the field stars are the small number statistics and also the unknown mass dependence of multiplets.
Therefore, the solar neighborhood sample  may not be representative of
the observations of Trumpler~19 because we cannot reach the end of the
main sequence  at  that  distance. 
Therefore,  given the  uncertainties, the
proportions  are comparable.   A  deeper multiplicity  census of  open
clusters such  as  Trumpler~19, which  represents a  controlled environment
with known physical properties, is essential for a comparison with the solar
neighborhood population.   This may be achieved  using future datasets
such as Gaia DR4 and LSST/Rubin in the optical, in combination with Roman
Space Telescope near-IR imaging.

In summary, we confirm previous results that there is a hefty sequence
of        unresolved       binaries        in       Trumpler        19
\citep{Jadhav2021,Donada2023,Jiang2024,Sheikh2025}. We also report the
discovery of a sequence of  unresolved triple and quadruple systems in
this  cluster, providing  spectroscopic  confirmation of  some of  the
multiple systems.
Within the uncertainties, the fraction of multiple stars in this cluster is comparable to that  of the solar neighborhood,  with the caveat that  the local
sample  probes  deeper  down  the main sequence.   This  fraction  of
unresolved triples and quadruples is also  higher than that of the
open cluster M~67, which has the same age and chemical composition.

\section{Conclusions}

To put it  in a broad context, this work  illustrates the relevance of
the  large surveys  for the  study of  multiplicity in  star clusters.
Gaia photometry and astrometry helped to secure cluster membership and
study the binary sequences present in a large number of star clusters.
In  addition,  the  VVVX  near-IR   photometry  helps  to  search  for
higher-order multiple  systems that lie  above the binary  sequence of
equal-mass stars.   The VVVX-GalCen spectroscopic survey allows us  to close
the loop,  confirming membership  for unresolved triple  and quadruple
systems in star clusters such as Trumpler~19.

Unresolved  triples and  quadruples have  been previously  detected in
some  young  open  clusters  and associations  with  ages of  $t<100$~Myr
\citep[e.g.,  ][]{Gonzalez2014,Tokovinin2023,Malofeeva2023}. However,
the main result of this work is  the proof that a population of triple
and quadruple  systems can manage to  survive in an aged  open cluster
for a long  time (over gigayears).  This leads to  many open questions
that need to be answered:

—  How  a large  number of  multiple systems  affects the  cluster
evolution. This scenario  was explored  by \citet{Kiseleva1996}  and
\citet{Leigh2013}. 

—  Why  Trumpler~19  has  more  unresolved  multiple systems than M~67, a cluster with a similar mass and age.

— Whether  this is a primordial  feature, or  the result  of dynamical
evolution  (i.e.,  M~67  managed   to  destroy  multiple  systems  more
efficiently than Trumpler~19).

— How the fraction of multiple stars depends on mass as we go down
the  main sequence.  In general, the fraction of high-order hierarchical systems seems to depend strongly on mass (Offner et al. 2023).
For Trumpler~19,  \citet{Jadhav2021}   found  an
increasing binary  fraction with  mass for more  massive stars  in the
upper  MS, but  it is  unknown if  this trend  continues to  the lower
MS. It would be interesting to  probe if the frequency of binaries and
triples remains  the same farther  down the main sequence,  for M-type
stars with  much lower masses.  This can be  tested in a  cluster such as
Trumpler~19 using deeper observations.

— What new tools we need to develop. For example, the new synthetic
CMD   codes   that   fit    the   cluster   parameters   such as   SIESTA
\citep[statistical   matching  between   real  and   synthetic  stellar
  populations;][]{Ferreira2024}   need  to   be  updated   to  include
non-resolved triples and quadruples, not just binaries.

— What the actual number of merged stars within this cluster is, and whether  the presence  of triple  and  higher-order  systems is important  (or
essential) for the formation of blue stragglers.

— What  the initial  fraction of  multiple stars was, given  that this  cluster has lost  most of its  mass, and how  this compares with the present value.

Our immediate follow-up plan is to  secure the detection of the triple
sequence in other suitable open clusters, such as NGC2437.  The potential to
extend further this work is promising, given by the pending release of
Gaia DR4, a  new version of the VIRAC3 catalog  from the VVVX survey
\citep[see ][]{Smith2025},
in combination with the
Roman   Space  Telescope   Galactic   Plane   Survey  and   LSST/Rubin
observations of the  Galactic plane, which will increase  the sample of
monitored clusters  with light  curves, down  to the  end of  the main sequence.
In  addition, some  of  the  wider systems  may  also  be resolved  by
follow-up observations with  adaptive optics systems from the ground,  and with the
Roman Space Telescope or the James Webb Space Telescope in space.

\begin{acknowledgements}

We gratefully acknowledge  the use of data from the  ESO Public Survey
program IDs 179.B-2002  and 198.B-2004 taken with  the VISTA telescope
and data  products from the Cambridge Astronomical Survey  Unit. 
D.M. is supported  by ANID Fondecyt Regular  No. 1220724.
R.K.S. is supported by CNPq/Brazil   through  projects 308298/2022-5 and 421034/2023-8.
C.O.O. is supported by the 2025 Postdoctoral Talent Attraction Competition for Research Centres and Institutes of Universidad Andres Bello, Project No. DI-07-25/ATP.
M.G.  and S.F. are  supported by ANID Fondecyt  Regular No. 1240755.  
J.A.-G.  is supported by  ANID Fondecyt Regular  No.   1201490.  
J.G.F-T  is supported by ANID Fondecyt Regular No. 1260371, by the Joint Committee ESO-Government of Chile under the agreement 2023 ORP 062/2023, and by the Doctoral Program in Artificial Intelligence, DISC-UCN.
 P.W.L. is supported by STFC grant ST/Y000846/1.  
 I.P. acknowledges support from ANID BECAS/DOCTORADO NACIONAL 21230761.
 C.C.,  B.D., and D.M. are supported by the ANID BASAL Center for  Astrophysics and Associated Tecnologies (CATA) project FB210003.  
 J.B.  is supported by ANID Fondecyt Regular  No.   1240249.   
 E.R.G. and A.L.  gratefully  acknowledge   the  ESO Fellowship program.  
 Z.G. is supported by ANID Fondecyt Iniciacion  No. 11260176.  
 We  also acknowledge support by the China-Chile Joint Research Fund (CCJRF No.2301) and the Chinese  Academy  of  Sciences  South  America  Center  for  Astronomy (CASSACA) Key Research Project E52H540301.
 T. P. is supported by the Argentinian  Secretaría de  Ciencia  y  Tecnología de  la Universidad  Nacional de  Córdoba (SECYT-UNC,  Res. 258/53).  
 R.K. is supported by ANID Fondecyt Regular grants No. 1261142 and 1240249.
This research has made use of the SIMBAD database, CDS, Strasbourg
Astronomical Observatory, France, and of data from the European Space Agency (ESA) mission \textit{Gaia}
(\url{http://www.cosmos.esa.int/gaia}), processed by the \textit{Gaia}
Data Processing and Analysis Consortium (DPAC, \url{http://www.cosmos.esa.int/web/gaia/dpac/consortium}).

\end{acknowledgements}

\makeatletter
\renewcommand\NAT@nmfmt[1]{\textsc{#1}}
\def\maxbibnames{2}  
\makeatother


\begin{thebibliography}{}
  
\bibitem[Alonso-Garc\'ia et al.(2026)]{AlonsoGarcia2026}
Alonso-Garc\'ia, J., Hempel, M., Saito, R.~K., et al.\ 2026, \aap, 706, A301

\bibitem[Brewer et al.(2016)]{Brewer2016}
Brewer, L.~N., Sandquist, E.~L., Mathieu, R.~D., et al.\ 2016, \aj, 151, 66

\bibitem[Cantat-Gaudin \& Anders(2020)]{CantatGaudin2020}
Cantat-Gaudin, T., \& Anders, F.\ 2020, \aap, 633, A99

\bibitem[Dias et al.(2021)]{Dias2021}
Dias, W.~S., Monteiro, H., Moitinho, A., et al.\ 2021, \mnras, 504, 356

\bibitem[Donada et al.(2023)]{Donada2023}
Donada, J., Anders, F., Jordi, C., et al.\ 2023, \aap, 675, A89

\bibitem[Duquennoy \& Mayor(1991)]{Duquennoy1991}
Duquennoy, A., \& Mayor, M.\ 1991, \aap, 248, 485

\bibitem[Ferreira et al.(2024)]{Ferreira2024}
Ferreira, B., Santos, J.~F.~C., Dias, B., et al.\ 2024, \mnras, 533, 4210

\bibitem[Freudling et al.(2024)]{Freudling2024}
Freudling, W., Zampieri, S., Coccato, L., et al.\ 2024, \aap, 681, A93

\bibitem[Gaia Collaboration (2023)]{Drimmel2023}
  Gaia Collaboration, Drimmel, R., Romero-G\'omez, M., et al.\ 2023, \aap, 674, A37

\bibitem[Gonzalez et al.(2014)]{Gonzalez2014}
Gonzalez, J.~F., Veramendi, M.~E., \& Cowley, C.~R.\ 2014, \mnras, 443, 1523

\bibitem[Hajdu et al.(2019)]{Hajdu2019}
Hajdu, T., Borkovits, T., Forg\'acs-Dajka, E., et al.\ 2019, \mnras, 485, 2562

\bibitem[Henry et al.(2018)]{Henry2018}
Henry, T.~J., Jao, W.-C., Winters, J.~G., et al.\ 2018, \aj, 155, 6, 265. doi:10.3847/1538-3881/aac262

\bibitem[Hunt et al.(2018)]{Hunt2023}
Hunt, E. L., \& Reffert, S.\ 2023, \aap, 673, A114

\bibitem[Hurley et al.(2005)]{Hurley2005}
Hurley, J.~R., Pols, O.~R., Aarseth, S.~J., \& Tout, C.~A.\ 2005, \mnras, 363, 293

\bibitem[Husser,  et al.(2013)]{Husser2013}
Husser, T.-O., Wende-von Berg, S., Dreizler, S., et al.\ 2013, \aap, 552, A6

\bibitem[Jadhav \& Roy(2021)]{Jadhav2021}
Jadhav, V.~V., \& Roy, K.\ 2021, \aj, 162, 264

\bibitem[Jiang et al.(2024)]{Jiang2024}
Jiang, Y., Zhong, J., Qin, S., et al.\ 2024, \apj, 971, 71

\bibitem[Kiseleva et al.(1996)]{Kiseleva1996}
Kiseleva, L., Aarseth, S., \& Eggleton, P.\ 1996, ASP, 90, 422

\bibitem[Leigh \& Geller(2013)]{Leigh2013}
Leigh, N.~W.~C., \& Geller, A.~M.\ 2013, \mnras, 432, 2474

\bibitem[Leiner et al.(2025)]{Leiner2025}
  Leiner, E.~M., Mathieu, R.~D., Geller, A.~M., et al.\ 2025, \apj, 980, 10

\bibitem[Malofeeva et al.(2023)]{Malofeeva2023}
Malofeeva, A.~A., Mikhnevich, V.~O., Carraro, G., \& Seleznev, A.~F.\ 2023, \aj, 165, 45

\bibitem[Raghavan et al.(2010)]{Raghavan2010}
Raghavan, D., McAlister, H. A., Henry, T. J., et al.\ 2010, \apjs, 190, 1

\bibitem[Offner et al.(2023)]{Offner2023}
Offner, S. S. R., Moe, M., Kratter, K. M., et al.\ 2023, ASPC, 534, 275O 

\bibitem[Reyle et al.(2021)]{Reyle2021}
Reyle, C., Jardine, K., Fouque, P., et al.\ 2021, \aap, 650, A201

\bibitem[Rappaport et al.(2024)]{Rappaport2024}
Rappaport, S.~A., Borkovits, T., Mitnyan, T., et al.\ 2024, \aap, 686, A27

\bibitem[Saito et al.(2025)]{Saito2024}
Saito, R.~K., Hempel, M., Alonso-Garc\'ia, J., et al.\ 2025, \aap, 689, A148

\bibitem[Sandquist et al.(2003)]{Sandquist2003}
  Sandquist, E.~L., Latham, D.~W., Shetrone, M.~D., \& Milone, A.~A.~E.\ 2003, \aj, 125, 810

\bibitem[Saydjari et al.(2023)]{Saydjari2023}
Saydjari, A.~K., Schlafly, E.~F., Lang, D., et al.\ 2023, \apjs, 264, 28

\bibitem[Sheikh et al.(2025)]{Sheikh2025}
Sheikh, A.~H., Medhi, B.~J., \& Sagar, R.\ 2025, \apj, 989, 16

\bibitem[Smith et al.(2025)]{Smith2025}
Smith, L.~H., Lucas, P.~W., Koposov, S.~E., et al.\ 2025, \mnras, 536, 3707

\bibitem[Tokovinin(2014)]{Tokovinin2014}
Tokovinin, A.\ 2014, \aj, 147, 87

\bibitem[Tokovinin(2023)]{Tokovinin2023}
Tokovinin, A.\ 2023, \aj, 165, 220

\bibitem[Torres et al.(2021)]{Torres2021}
Torres, G., Vanderburg, A., Curtis, J.~L., et al.\ 2021, \apj, 921, 133

\bibitem[Vrijmoet et al.(2026)]{Vrijmoet2026}
  Vrijmoet, E.~H., Tokovinin, A., Henry, T., et al.\ 2026, \aj, 171, 186

\end{thebibliography}
\end{document}